\documentclass[manuscript]{aastex}
\def\H2{\hbox{H$_{2}$}}
\def\deg{$^{\circ}$}

\def\Msun{\hbox{\it M$_\odot$}}
\def\one{{\,\sc i}}
\def\two{{\,\sc ii}}

\begin{document}

\title{New Near-Infrared Imaging and Spectroscopy of NGC 2071-IR}

\shortauthors{Walther \& Geballe}

\shorttitle{Background IR Sources in the Galactic center}

\author{D. M. Walther\altaffilmark{1}}

\author{T. R. Geballe\altaffilmark{1}}

\altaffiltext{1}{Gemini Observatory, 670 N. A'ohoku Place, Hilo, HI 96720; tgeballe@gemini.edu}

\begin{abstract}

We present high-resolution images of NGC 2071-IR in the $J$, $H$, and $K$ bands and in the emission at 2.12 $\mu$m of the $v=1-0~S$(1) line of molecular hydrogen. We also present moderate-resolution $K$-band spectra of two young stellar objects, IRS 1 and IRS 3,  within NGC 2071-IR, that are candidate sources of one or more of the outflows observed in the region. Two of the eight originally identified infrared point sources  in NGC 2071-IR are binaries, and we identify two new sources, one coincident with the radio source Very Large Array-1 and highly reddened.   The  \H2\  $Q$(3)/$S$(1) line intensity ratios at IRS 1 and IRS 3 yield high and very high extinctions, respectively, to them, as is implied by their near-infrared colors and K-band continuum slopes. The spectra also reveal the presence of hot, dense circumstellar molecular gas in each, suggesting that both are strong candidates for having energetic molecular outflows. We agree with a previous suggestion that IRS 1 is the likely source of an east$-$west oriented outflow and conclude that this outflow is probably largely out of the plane of the sky. We also conclude that if IRS 3 is the source of the large scale northeast$-$southwestoutflow, as has been previously suggested, its jet/wind must precess in order to explain the angular width of that outflow.  We discuss the natures of each of the point sources and their probable contributions, if any, to the complex morphology of the \H2\ line emission. 

\end{abstract}


\section{INTRODUCTION}

Four arcminutes north of the optical reflection nebula NGC 2071 lies an extended bipolar molecular outflow, first observed in the CO $J$=1-0 line by \citet{bal82}. The extent and northeast$-$southwest (NE-SW) orientation of the outflow have been well characterized in the $1-0$ and $2-1$ CO lines by \citet{sne84} and \citet{mor89}.   At the center of the outflow is NGC 2071-IR \citep{per81}, an intermediate-mass star-forming region, containing a cluster of at least eight near-IR sources, IRS 1$-$8 \citep[][hereafter W93]{wal93}, presumably young main-sequence stars or protostars.  None of them has a visible counterpart, thus all of them must lie within or behind the NGC 2071 molecular cloud.  \citet{sne86} found that each of the three brightest sources in the \citet{per81} 10 $\mu$m map, IRS 1, IRS 2, and IRS 3, is associated with a compact radio source. 

A high-resolution spectrum of the 1$-$0 $S$(1) line of molecular hydrogen (\H2) at a position within NGC 2071-IR obtained by \citet{per81} revealed a wide velocity profile similar to those previously seen in the core of the Orion Molecular Cloud \citep{nad79}, indicating that the emitting \H2\ is shock-excited, a conclusion subsequently confirmed by \citet{bur89}  The emission in this line has been mapped by \citet{lan86} at $15"- 20''$ resolution, by \cite{bur89} using a $19''$ beam, by \citet{gar90} using a camera with 1\farcs2 pixels, by \citet{asp93} using a camera with 0\farcs6 pixels, and most recently, by \citet[][hereafter E00]{eis00} using a camera with 0\farcs8 pixels. \citet{bur89} found that the shocked \H2\ in the NE lobe is blue-shifted relative to the \H2\ in the SW lobe, indicating that the direction of the NE outflow, while largely in the plane of the sky, is also toward the front surface of the molecular cloud. Both IRS 1 and IRS 3 are located at a minimum in the \H2\ line emission strength between the NE and SW outflows, whereas IRS 2 is displaced to the northeast of this minimum.

Although the large-scale morphology of the \H2\ line emission crudely resembles that of the CO millimeter line emission, and is suggestive of a single dominant outflow, the \H2\ line emission in the central part of NGC 2071-IR, is spatially complex. Some investigators (W93, E00) have suggested that other smaller-scale outflows may be present in that region. Based on the orientations of centimeter-wavelength continuum emission and H$_{2}$O maser spots observed at IRS 1 and IRS 3 by \citet[][hereafter T98]{tor98}, who identified both objects as outflow sources,  E00 has suggested that IRS 3, rather than IRS 1,  is the source of the large-scale NE$-$SW outflow.  E00 also similarly argued that IRS 1 is the source of a second, bright but much smaller scale (on the plane of the sky) flow oriented approximately east$-$west (EW). (Note that an EW orientation of the radio continuum emission at  IRS 1 had previously been observed by \citet{smi94}.) Both W93 and E00 also have suggested that IRS 7, which is located well to the west of the NE$-$SW outflow, is the source of another outflow. 

Subsequent radio observations  at the Very Large Array (VLA) by \citet[][hereafter TRR09]{tri09} have resolved both IRS 1 and IRS 3 into three components separated from one another by a few tenths of an arcsecond. They interpret the outer components as condensations of ionized gas being ejected by the central objects. \citet[][hereafter CG12]{car12}  also have detected radio jets from these two objects. TRR09 and \citet{set02} mapped the H$_{2}$O masers  with the VLA and the Very Long Baseline Array, respectively, and interpret those data as evidence for disks surrounding each central object. TRR09 and CG12 also detected an additional radio source, VLA-1, located in between IRS 1 and IRS 3, for which no infrared counterpart was known.  Skinner et al., 2009 (hereafter S09) found X-ray emission at or close to a number of the IRS objects, including IRS 1, IRS 3 and VLA-1.

\citet[][hereafter W91]{wal91} found Br~$\gamma$ (2.17 $\mu$m) line emission and CO overtone (2.3$-$2.4 $\mu$m) band emission at IRS 1.  Their spectrum of IRS 3 also appears to show CO in emission, but the signal-to-noise ratio is low. They also found CO overtone band emission in IRS 7.   Both Br~$\gamma$  emission and CO emission can be signs of pre-main sequence accretion and/or mass loss.  The very high infrared polarizations of IRS 1 and IRS 3 found by W93 and \citet[][hereafter T07]{tam07} indicate that the radiation from each does not reach us directly from the central protostar, but is scattered by dust particles close to these sources.

In view of the proximity \citep[$d$ = 390 pc][]{ant82} of this intermediate-mass star forming region, the much-improved infrared sensitivities and higher angular resolution available since previously published infrared imaging observations, and the morphological complexity of the NGC 2071-IR region, we have obtained new near-infrared images of NGC 2071-IR. We also have obtained new $K$-band spectra of IRS 1 and IRS 3, for which the previous published spectra date from 1990. The work described here is part of our continuing project to link the pointlike IR sources in NGC 2071-IR sources to the large-scale structures seen in \H2, in order to arrive at an improved understanding of the spatial morphology of the \H2\ line emission.

\section{OBSERVATIONS AND DATA REDUCTION}

High-resolution images of NGC 2071-IR in broadband $J$, $H$, $K$ and narrowband \H2\ 1$-$0 $S$(1) 2.12 $\mu$m line and 2.09 $\mu$m continuum filters (with passband FWHMs of 1.31\%\ and 1.23\%, respectively) were obtained at the Frederick C. Gillett Gemini North Telescope on Maunakea, using Gemini's facility Near-infrared Imager  NIRI \citep{hod03} on the night of UT 2017 December 26, for program GN-2017B-DD-11. The observations utilized the NIRI $f$/6 camera, which provides a plate scale of 0\farcs117 / pixel and a field of view of $120'' \times\ 120''.$

The telescope was centered on NGC 2071 IRS 1. The sky positions, observed nearly simultaneously with NGC 2071-IR, were centered 3\farcm5 to the northwest. Both NGC 2071-IR and sky were observed using 6-point dither patterns. The sky was photometric and the seeing was excellent, with FWHMs of approximately 0\farcs40 in the $J$ band and 0\farcs35 in the $H$ and $K$ bands. Total on-source and sky exposure times were each 90, 96, and 108 s at $J$, $H$, and $K$, respectively, and 360 s in both of the narrowband filters. 

The data were processed using the Gemini IRAF and Python (aka Gemini PyRAF) packages. Standard techniques of linearizing, cleaning, flat-fielding, dark-subtraction, creating median sky frames, and subtracting them from the on-source reduced frames were employed. The final sky-subtracted \H2\ image was continuum-subtracted using the sky-subtracted 2.09 $\mu$m continuum image. Prior to subtraction a slight shift was made to align the point sources in the two images.  A small scaling factor was applied to the 2.09 $\mu$m image prior to subtraction from the 2.12 $\mu$m image, so that stars located well off strong \H2\ emission were approximately removed. Note that this does not ensure that the point sources within NGC 2071-IR are completely removed in the continuum-subtracted image, as their continua have different slopes than those of the stars used for scaling the images in the two filters.

 All NIRI images were airmass-corrected using the nominal atmospheric extinctions for Maunakea of $J$ = 0.015, $H$ = 0.059 and $K$ = 0.033 mag per airmass.  Software aperture photometry was performed on the $JHK$ images, resulting in the photometry presented in Table 1. Photometry for most of the point sources is in 1\farcs40 diameter apertures, with different annuli surrounding the different apertures to subtract extended emission. Most objects were detected at very high signal-to-noise ratios, but the accuracy of the photometry is dependent on the uniformity of the extended emission near each object, which is difficult to estimate.  

$K$-band spectra of IRS 1 and IRS 3 were obtained as part of program GN-2018B-FT-107 on UT 2018 November 10 in $K$-band seeing of 0\farcs67  (FWHM) and very thin cirrus, again using the Gemini North telescope, but in this case its near-infrared spectrograph, GNIRS \citep{eli06}. The spectrograph was configured with its short blue camera (0\farcs15 pixels), 32 l/mm grating, and a 0\farcs45 wide slit, which produce a resolving power of 1200, corresponding to a velocity resolution of 250 km s$^{-1}$. The slit was oriented at position angle $160^{\circ}$ in order to simultaneously obtain spectra of these objects, which are separated by 5\farcs6, and the telescope was nodded 3\farcs0 along the slit. The total exposure time was six minutes. Standard data reduction techniques were employed using standard IRAF and Figaro commands. The spectrum of a nearby early A-type star, HIP 30692, obtained immediately after the spectra of IRS 1 and IRS 3, was employed both to remove telluric absorption lines and serve as a flux-calibrator.

\begin{table}[http]
\renewcommand{\arraystretch}{.8}
\caption{Point-source Positions and Photometry $^a$} 
\label{sources}
\begin{center}
\scriptsize
\begin{tabular}{lccccccc}
\hline
Source& RA Offset$^b$ & Dec Offset$^b$& $J$ & $H$ & $K$ & $J-H$ & $H-K$ \\
\hline
IRS 1$^c$ & 0 & 0 & 16.60 & 13.27 & 10.99 & 3.33 & 2.29 \\
IRS 2A$^d$ & 8.86 & 7.47 & 17.5 & 14.3 & 12.0 & 3.2 & 2.3 \\
IRS 2B$^d$ & 10.04 & 7.30 & 16.9 & 15.2 & 14.5 & 1.7 & 0.7 \\
IRS 3 & -1.97 & 5.28 & $>$21.4 & $>$21.0 & 15.20 & ... & $>$5.8 \\
IRS 4 & 5.36 & 18.71 & 15.62 & 12.97 & 11.26 & 2.68 & 1.71 \\
IRS 6A & 14.30 & 27.57 & 14.07 & 11.77 & 10.35 & 2.30 & 1.42 \\
IRS 6B & 12.55 & 28.39 & 14.38 & 12.49 & 11.56 & 1.89 & 0.93 \\
IRS 7 & -11.55 & 27.67 & $>$21.4& 15.79 & 12.04 & $>$5.5 & 3.75 \\
IRS 8 & -6.87 & -4.73 & $>$21.4 & 17.70 & 15.28 & $>$3.7 & 2.42 \\
IRS 61 & 10.34 & 13.58 & 18.19 & 16.85 & 16.03 & 1.34 & 0.82 \\
VLA-1$^e$ &{\bf 0.32} & {\bf 2.39}$^e$ & $>$21.4 & $>21.0$ & 16.81 & ... & $>$3.1 \\
\hline
\end{tabular}
\end{center}
\scriptsize
$^a$In magnitudes, on the Vega system, in a 1\farcs40 diameter aperture; magnitude uncertainties $\sim$0.05 except where noted.\\
$^b$In $K$-band images, from IRS 1 (2MASS RA = 05:47:4.78, Dec = 0:21:42.8 (2000)); uncertainty $< $ 0\farcs05.\\
$^c$Measured in a 2\farcs34 diameter aperture.\\
$^d$Measured in a 0\farcs70 diameter aperture; uncertainties $\sim$0.1 mag to due to proximity of companion and strong and non-uniform extended emission.\\
$^e$Incorrect values in ApJ paper (2019, ApJ, 875, 153).\\

\end{table}

\section{RESULTS }

\begin{figure}  
\centerline{
\resizebox{0.95\textwidth}{!}{\includegraphics[angle=0]{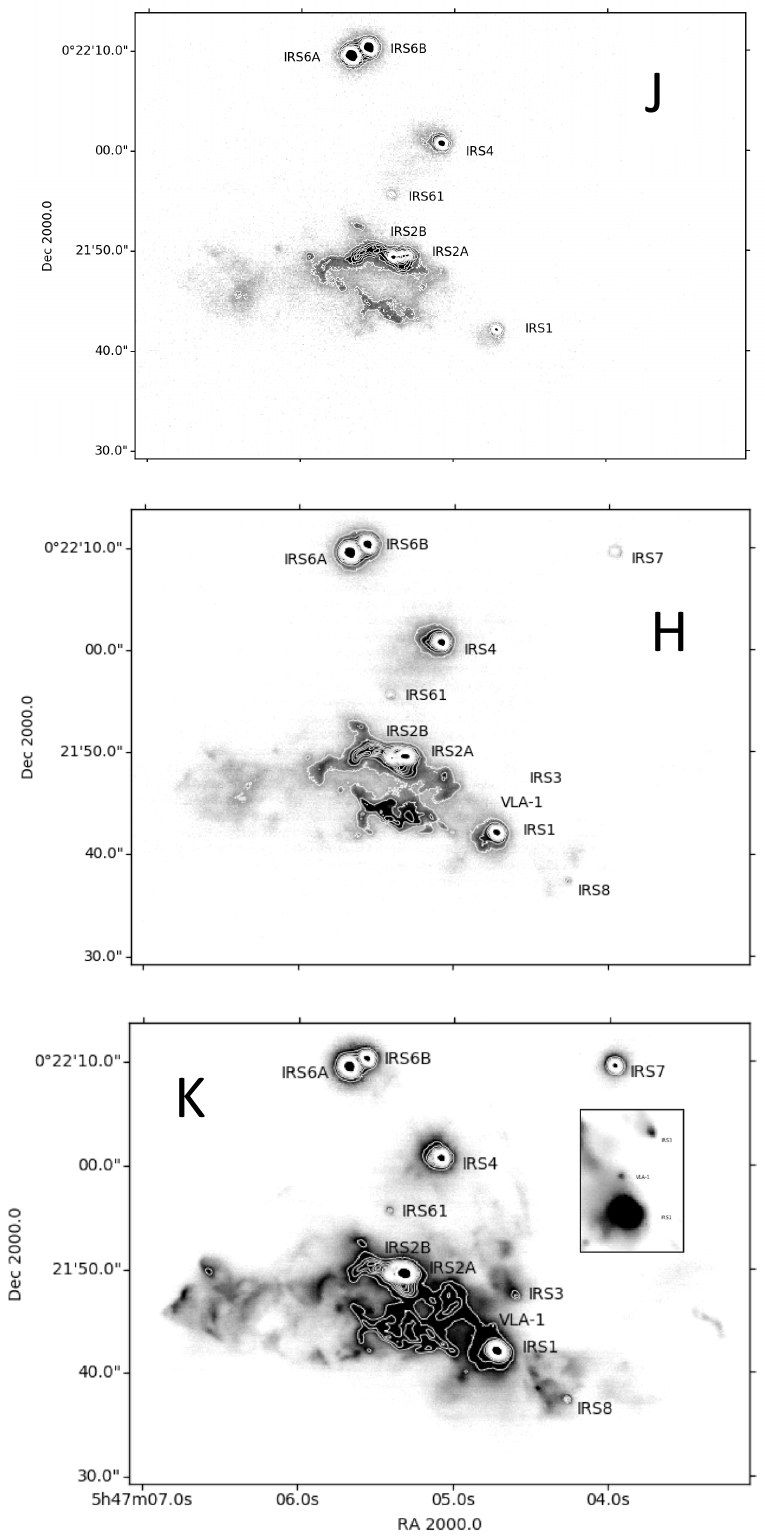}}}
\caption{ Images of the NGC 2071-IR region at $J$, $H$, and $K$, with IRS stellar sources marked.  Contour lines are evenly spaced in intensity. The inset in the $K$-band image is an expanded view of the region containing IRS 1, IRS 3, and VLA-1. }
\label{fig1}
\end{figure}

\subsection{Broadband Images}

The $J$, $H$ and $K$ images shown in Figure 1 reveal point-like sources in NGC 2071-IR as well as extended emission. Eleven unresolved sources are detected in the K-band image and are single stars or unresolved multiple-star systems. Not all of them are detected at $J$ and $H$. Table 1 gives their offsets from IRS 1, $JHK$ photometry, and colors.  All have colors redder than main-sequence stars and thus are obscured by interstellar dust within NGC 2071-IR. The most highly reddened of these are probably also obscured by their own dusty circumstellar material.  

Two embedded sources, IRS 2 and IRS 6, are newly discovered infrared binaries and hereafter referred to as IRS 2A and IRS 2B and IRS 6A and IRS 6B. IRS 2 binarity was previously observed at 3 mm by CG12, but its orientation differs somewhat from the infrared binary (see the discussion in Section 4.2). In each case the A source is considerably brighter than the B source at $H$ and $K$. Figure 2 contains contour plots of the more compact binary, IRS 2 at $J$, $H$, and $K$.  IRS 61 is a newly discovered source (we follow the IRS numbering convention of W93). Its $JHK$ colors are not as red as any of the other objects, but they are redder than an unobscured M dwarf, and thus it is not a foreground star. It could be either a star within but near the front surface of the NGC 2071 molecular cloud or a  mid L-type brown dwarf at a distance of a few tens of parcsecs.  A faint infrared counterpart to the radio source VLA-1 is also detected in the $K$ band (see the inset to Figure 1, bottom panel). Three IRS objects, IRS 5, IRS 5a, and IRS 8a, which were identified by W93 as pointlike sources of \H2\ line emission, are compact, but spatially resolved. We confirm that they do not coincide with peaks in the continuum and thus do not contain stars.

The new photometry is not easy to compare with the previous photometry by W93, because of the widely different apertures used, but is crudely consistent in that almost all of the objects are fainter in the much smaller apertures of the current observations.  The source IRS 7 is the lone exception; its $K$ magnitude is currently 30\%\ brighter than in 1991, whereas at $H$ it is 20\%\ fainter. These differences suggest a significant change in the intrinsic spectrum and possibly a small decrease in the extinction.

\begin{figure}  
\centerline{
\resizebox{0.8\textwidth}{!}{\includegraphics[angle=0]{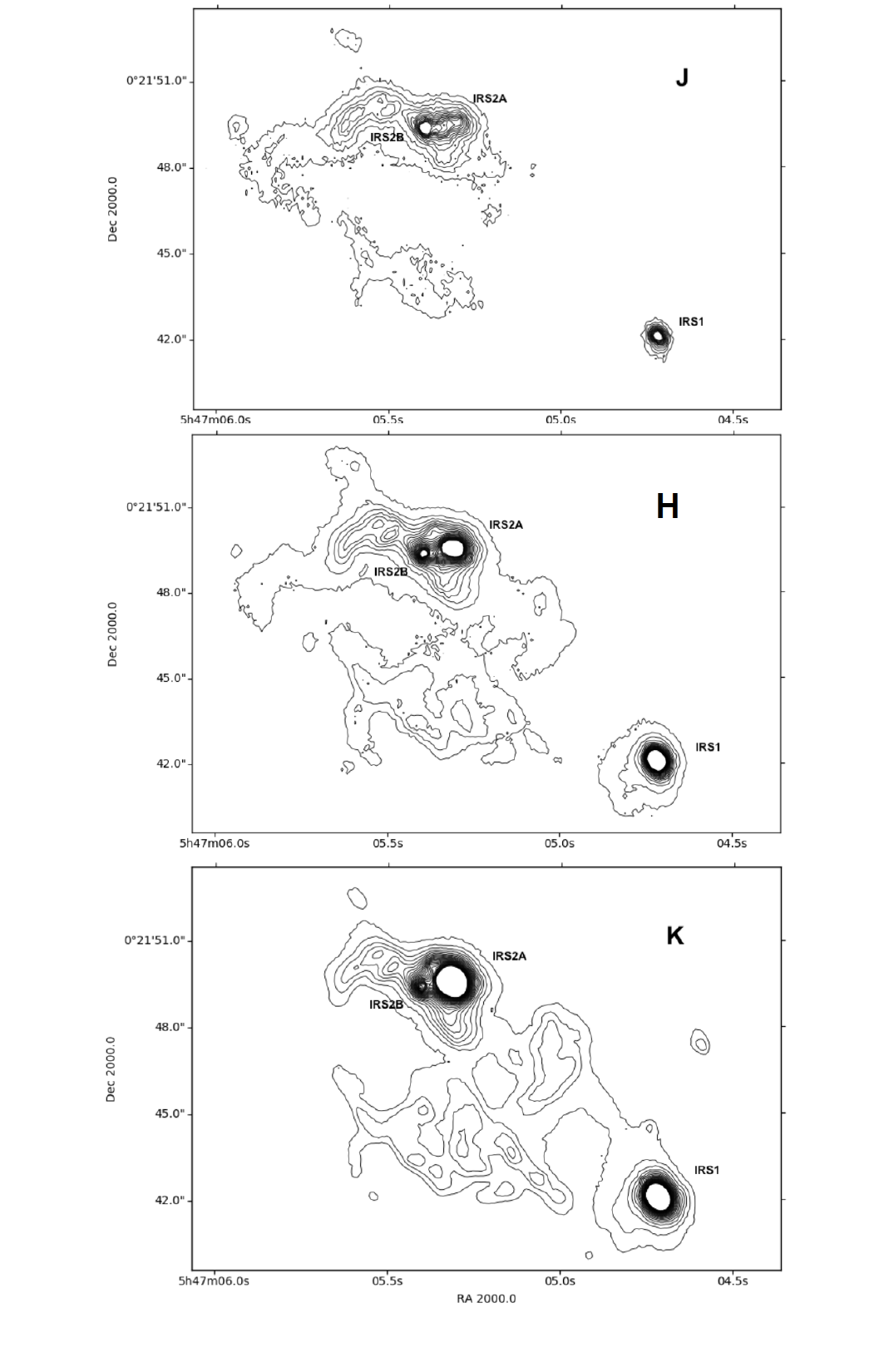}}}
\caption{$JHK$ contour maps of the the IRS 1 $-$ IRS 2 region.  The contour lines are evenly spaced intensity levels.}
\label{fig2}
\end{figure}

\subsection{Molecular Hydrogen Image}

\begin{figure}  
\centerline{
\resizebox{1.5\textwidth}{!}{\includegraphics[angle=0]{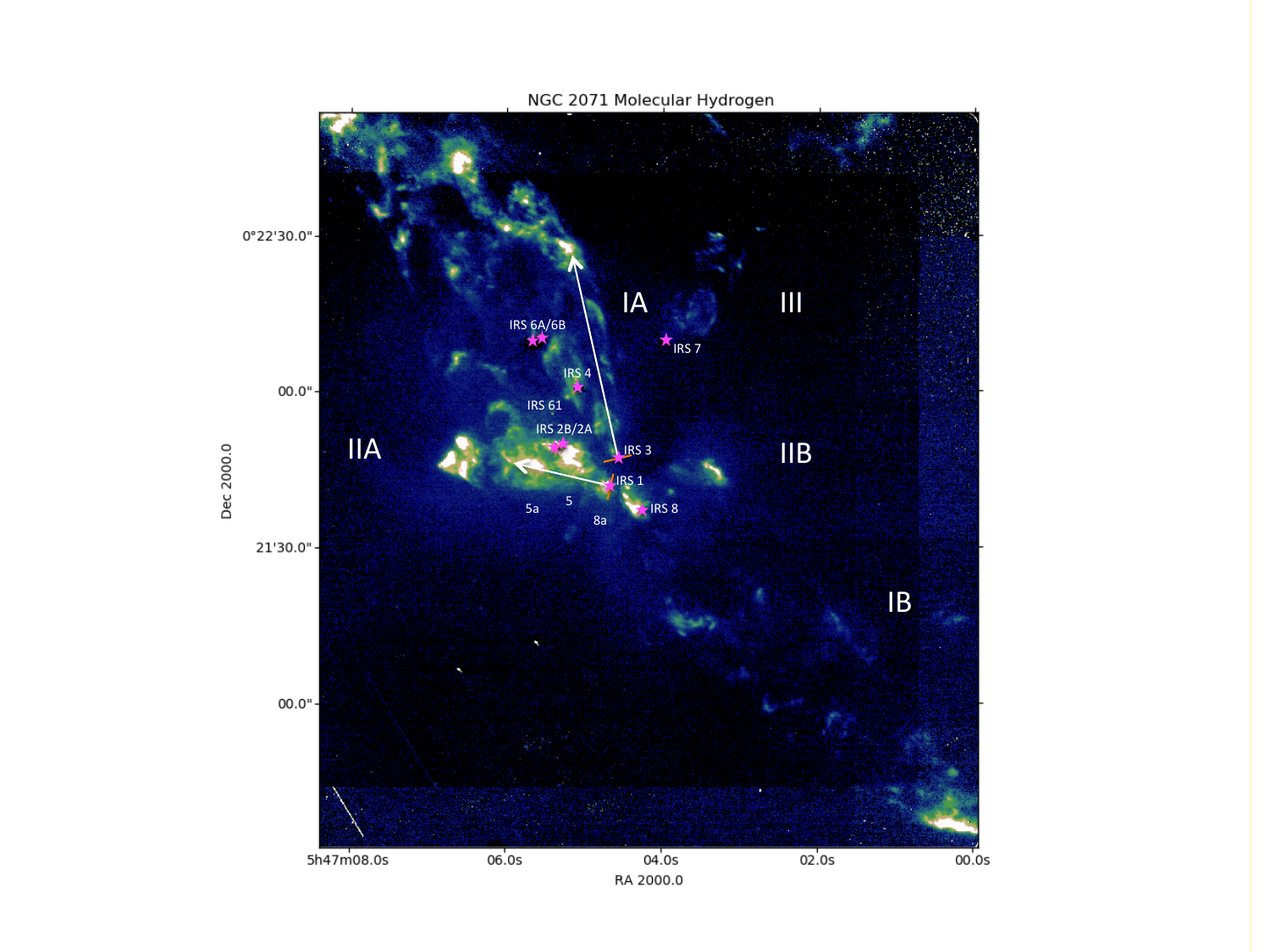}}}
\caption{Continuum-subtracted image of NGC 2071-IR in the 1$-$0 $S$(1) line of \H2. The positions of known (proto)stars are marked, along with the outflow regions IA-III designated by E00. Arrows denote directions of the E-pointing and NNE-pointing ionized jets from IRS 1 and IRS 3, respectively, as determined from radio continuum measurements by T98, S02, TRR09, and CG12}
\label{fig3}
\end{figure}

\begin{figure}  
\centerline{
\resizebox{1.4\textwidth}{!}{\includegraphics[angle=0]{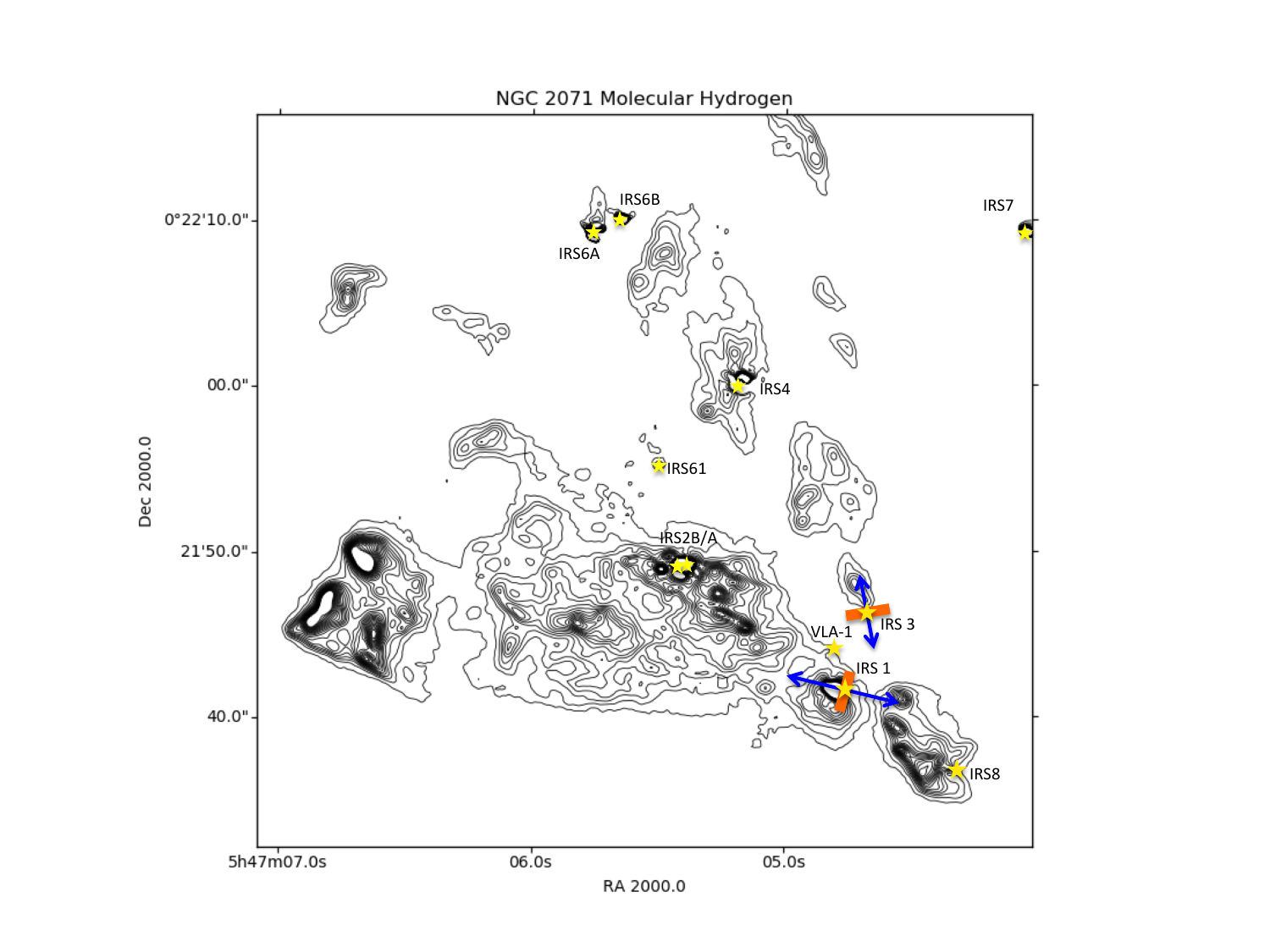}}}
\caption{Contour map of \H2\ $1-0$ $S$(1) line emission in the central region of Figure 3. The locations of IRS sources are denoted by yellow stars. Also shown are the orientations of the disks of IRS 1 and IRS 3 (orange rectangles), as determined from observations of H$_2$O maser spots by T98, and TRR09,  and their ionized jets (blue arrows), as determined from radio continuum measurements by T98, S02, TRR09, and CG12.}
\label{fig4}
\end{figure}

Figure 3 is a continuum-subtracted image of the \H2\ 1$-$0 $S$ (1) line  at 0\farcs35 resolution, covering $\sim$60\%\ of the full extent of NE$-$SW-oriented line emission (designated IA and IB in the figure, following E00) detected by earlier observers at lower angular resolution and imaged in full by E00 using a camera with 0\farcs81 pixels in unspecified seeing conditions. Figure 4 is a more detailed view, in contours of intensity, of the central $45''$. Outside of the central $30''$ many of the brightest emitting regions are located on the outer edges of the NE and SW lobes, suggesting that they are regions where an outflowing wind from a star is colliding with and shocking the surrounding cloud. Although these edges appear to be sharp, especially in the NE lobe (IA), the non-uniformity in surface brightness is remarkably great. Both IRS 1 and IRS 3 lie between these two emission lobes, and as noted by several previous investigators, are possible sources of this wind.  Source VLA-1 also is located between the lobes.   

Considerable \H2\ line emission is present in Figure 3 outside of IA and IB. Much of this was described by E00, but the present higher-resolution images reveal additional detail.  In particular, the morphology between those two lobes is highly complex, especially to the east. Knots of bright \H2\ line emission are present to the east of IRS 1, VLA-1, and IRS~3, extending as far as $40''$ from them, and are collectively called Outflow IIA by E00.  Approximately $15''$ to the west of these sources a bright arc of emission is present within a larger region of lower surface brightness, which E00 labeled IIB. Lying roughly between IRS 1 and IRS 8 is smaller region of bright line emission, which might be part of either Outflow IIA or Outflow IB. Finally, to the northwest of IRS 7 are several regions of mostly lower surface brightness line emission (Outflow III); the most distant of these is approximately one arcminute from IRS 7.

Many of the regions of bright line emission are embedded in lower surface emission, some of which could be scattered line radiation, an interpretation that would be consistent with the detections by W93 and T07 of high polarization in those regions.  In the immediate vicinities of IRS 1 and IRS 3 the \H2\ line emission is faint.  W93 and T07 have reported high polarizations of the continua of IRS 1 and IRS 3, which they suggest are due to scattering of their continua radiation by dusty disks.  These indicate that the line emission originating from and near each source is heavily obscured.

\begin{figure}  
\centerline{
\resizebox{0.65\textwidth}{!}{\includegraphics[angle=0]{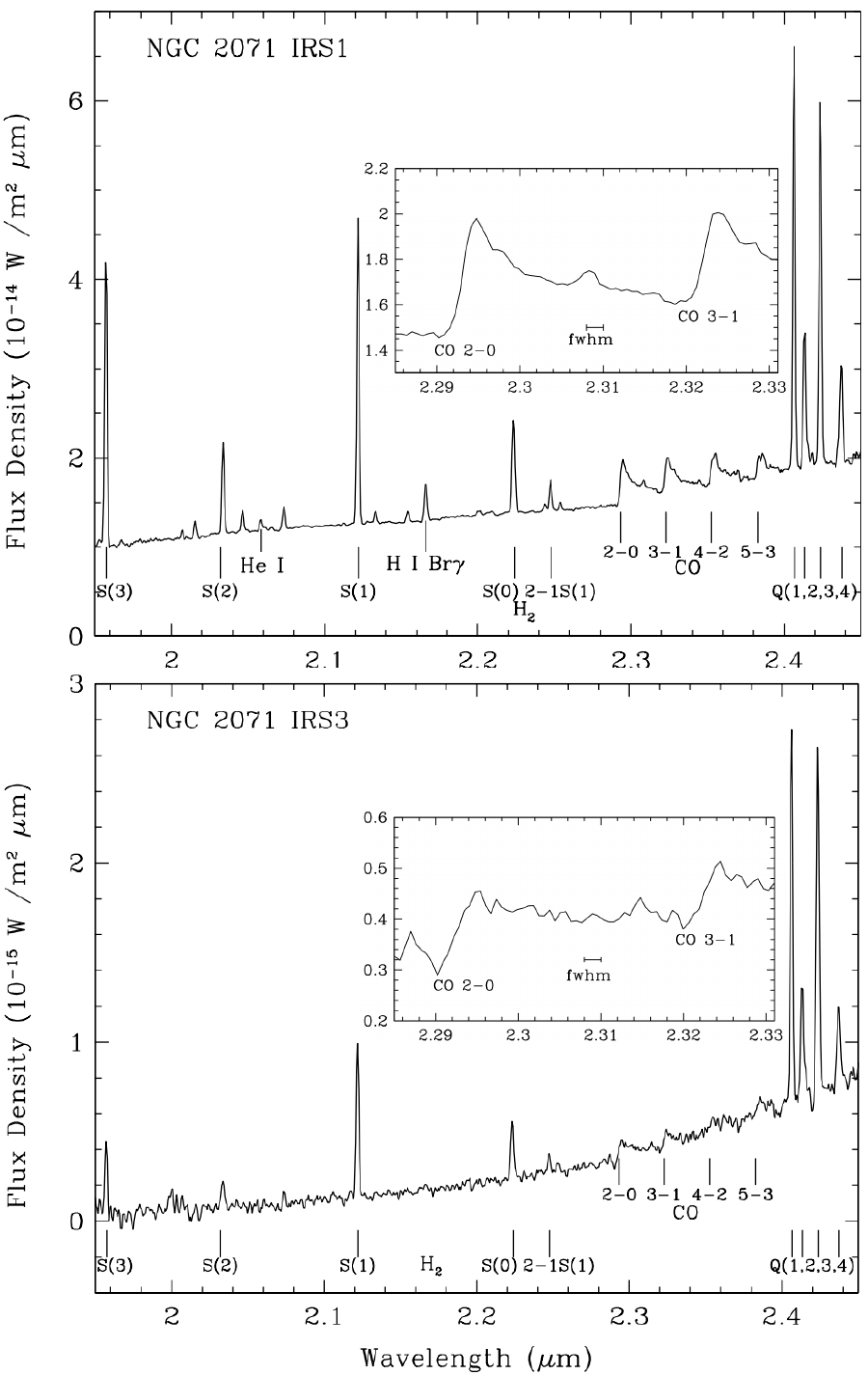}}}
\caption{Upper panel:  $K$-band spectrum of IRS 1 with the wavelengths of Br $\gamma$, the first four CO band heads, and the strongest lines of \H2\ (all but one from the $v = 1-0$ band) indicated by vertical lines. The unmarked lines are due to \H2, Fe\one, and [Fe\two] (see Table 2). The inset shows the details of the spectrum around the CO $2-0$ and $3-1$ band heads. Lower panel:  $K$-band spectrum of IRS 3 with all detected lines labeled, and inset showing details of the CO $2-0$ and $3-1$ bands. 
}
\label{fig5}
\end{figure}

\begin{table}[http]
\renewcommand{\arraystretch}{.8}
\caption{Emission lines in IRS 1 and IRS 3}
\label{linelist}
\begin{center}
\scriptsize
\begin{tabular}{cccccc}
\hline
Obs. Wavel. & Obs Wavel. & Lab. Wavel.$^a$ & Line ID & IRS 1 Flux & IRS 3 Flux \\
(vac. $\mu$m) & (vac. $\mu$m) & (vac. $\mu$m) & & & \\
IRS 1 & IRS 3 & & & 10$^{-17}$ W m$^{-2}$ & 10$^{-17}$ W m$^{-2}$ \\
\hline
1.9572 & 1.9571 & 1.9576 & \H2\ 1$-$0 $S$(3) & 6.26 & 0.072 \\
1.9673 & ... & 1.9675 & [Fe\two] & 0.14 & ... \\
2.0071 & ... & 2.0072 & [Fe\two] & 0.14 & ... \\
2.0155 & ... & 2.0157 & [Fe\two] & 0.53 & ... \\
2.0337 & 2.0336 & 2.0338 & \H2\ 1-0 $S$(2) & 2.00 & 0.036 \\
2.0464 & ... & 2.0466 &  [Fe\two] & 0.49 & ... \\
2.0584 & ... & 2.0587 & He\one\ 2P$-$2S & 0.17 & ... \\
2.0652 & ... & 2.0656 & \H2\ $3-2$ $S$(5) & 0.17 & ... \\
2.0735 & 2.0737 & 2.0735 & \H2\ 2$-$1 $S$(3) & 0.49 & 0.013 \\
2.1124 & ... & 2.1126 & He\one\ 3S$-$3P & 0.03 & ... \\
2.1218 & 2.1218 & 2.1218 & \H2\ 1$-$0 $S$(1) & 6.72 & 0.176 \\
2.1276 & ... & 2.1280 & \H2\ 3$-$2 $S$(4) & 0.03 & ... \\
2.1332 & ... & 2.1334 & [Fe\two] & 0.24 & ... \\
2.1542 & ... & 2.1542 & \H2\ 1$-$0 $S$(2) & 0.26 & ... \\
2.1661 & 2.1665 ? & 2.1661 & H\one\ 7$-$4 & 1.02 & 0.004 \\
2.2010 & ... & 2.2014 & \H2\ 3$-$2 $S$(3) & 0.12 & ... \\
2.2232 & 2.2233 & 2.2233 & \H2\ 1$-$0 $S$(0) & 2.23 & 0.071 \\
2.2440 & ... & 2.2241 & [Fe\two] & 0.13 & ... \\
2.2476 & 2.2475 & 2.2477 & \H2\ 2$-$1 $S$(1) & 0.65 & 0.019 \\
2.2538 & ... & 2.2541 & [Fe\two] & 0.16 & ... \\
2.3083 & ... & 2.3085 & Fe\one\ ? & 0.16 & ... \\
2.4064 & 2.4063 & 2.4066 & \H2\ 1-0 $Q$(1) & 9.22 & 0.419 \\
2.4134 & 2.4133 & 2.4134 & \H2\ 1-0 $Q$(2) & 3.45 & 0.140 \\
2.4237 & 2.4236 & 2.4237 & \H2\ 1-0 $Q$(3) & 8.44 & 0.408 \\
2.4372 & 2.4372 & 2.4375 & \H2\ 1-0 $Q$(4) & 2.85 & 0.118 \\
\hline
\end{tabular}
\end{center}
$^a$ Fe wavelengths from The Atomic Linelist v2.04, http://www.pa.uky.edu/$\sim$peter/atomic/

\end{table}

\subsection{$K$-band Spectroscopy}

The $K$-band spectra of IRS 1 and IRS 3 are shown in Figure 5 and the detected lines in each spectrum are listed in Table 2.  When convolved to the low resolution of the spectra in W91 they reveal no large changes in the spectra between 1990 and 2018.  

In many respects, the spectra of IRS 1 and IRS 3 are similar, with the line emission dominated by $1-0$ $S$- and $Q$-branch transitions of \H2\, and emission in the first four CO overtone bands, all superimposed on steeply rising continua. In IRS 1 many weaker lines of \H2\ are also detected; in IRS~3 the much lower signal-to-noise ratio precludes their detection. The spectrum of IRS~1 also contains some weak lines not due to \H2. We identify seven of them as due to forbidden Fe\two, based both on accurate wavelength agreement with laboratory wavelengths and the presence of many of them in the spectra of evolved  emission line stars \citep{ham94a,fig98,geb00}. One line at  2.308 $\mu$m is tentatively identified as due to Fe\one; this identification is based only on the excellent match to the laboratory wavelength and the presence  in the spectrum of many lines of [Fe\two].   To our knowledge the forbidden iron lines in the $K$ band had not previously been identified in the spectrum of any protostar. However, a few of the spectra of Class I protostars published by \citet{con10} appear to contain one or both of the strongest of these lines, at 2.0157 $\mu$m and 2.0466 $\mu$m, although that is not mentioned in their paper. Forbidden Fe\two\ lines at 1.53 $\mu$m and 1.64 $\mu$m have been found in the spectra of several low-intermediate-mass protostars by \citet{ham94b}, who deduced that they arise in ionized winds with densities $\sim$10$^{4}$ cm$^{-3}$. These two lines are  present in many of the spectra in \citet{con10}, and are much stronger than the $K$-band [Fe\two] lines. Further analysis and discussion of the $K$-band lines is outside the scope of this paper.

Examination of raw spectral frames shows that at both IRS 1 and IRS 3 there are compact regions of bright \H2\ line emission whose peaks coincide closely (within 0\farcs15) to the peaks in the continua of these objects.  These regions extend for about 2$''$ at IRS 1 and about 1$''$ at IRS 3. Thus, for the most part the \H2\ lines in the reduced spectra of IRS 1 and IRS 3 in Figure 5 arise in the vicinities of these objects rather than being the result of gradients in the extended line emission. The weak extended line emission should be at least partly removed by the 3$''$ nod and should be a minor contributor to the spectra in Figure 5. 

Some differences between the two spectra are not explainable by the large difference in signal-to-noise ratios. The continuum of IRS 3 is much redder than that of IRS 1, indicating that is much more deeply embedded in its dusty cocoon than IRS 1 is. This also is deduced  from the $JHK$ photometry. This is borne out by estimates of the extinction to the \H2\ line-emitting regions of each source from the ratio of the fluxes of the 1$-$0 $Q$(3) and $S$(1) lines, which arise from the same upper level. Using the procedure described in \citet{pik16}, and assuming, as they did, a $-1.7$ power-law dependency of the extinction on wavelength \citep{ind05}, we derive $A$$_K$ and $A$$_V$ of 3.0 mag and 37 mag, respectively, for IRS 1 and 6.1 mag and 76 mag, respectively, for IRS 3. To our knowledge the latter value of  $A$$_V$ is the largest ever determined using this technique. 

Another significant difference between the two spectra is in the sharpness of the CO band heads, as seen in the insets in Figure 5. Although the signal-to-noise ratio of the spectrum of IRS 3 is not high, it is clear that the widths of its band heads (from continuum at leading edge to peak) are nearly twice the instrumental resolution of 250 km s$^{-1}$, implying that the intrinsic velocity widths of the overtone CO lines in IRS 3 are $\sim$250 km s$^{-1}$.  In IRS 1 the band heads are unresolved, indicating that its intrinsic CO line widths are $<$ 125 km s$^{-1}$.  In contrast to the difference for CO lines, the \H2\ lines in both objects are unresolved. The high signal-to-noise profiles of the $1-0$ $S$(1) line in each imply \H2\ FWHMs $<$ 100 km s$^{-1}$.  In both cases the overtone emission from vibrational levels as high as $v=5$ must originate in hot and dense ($T \sim\ 3,000$ K, $n \gtrsim 10^{10}$ cm$^{-3}$ ) gas, located just outside of  the protostellar photosphere.

\section{DISCUSSION}

Our long-term goal is to combine infrared imaging and spectroscopy with radio and millimeter data to better understand the morphology of the \H2\ line emission in NGC 2071-IR and to identify the protostar(s) responsible for different regions of \H2\ line emission. The spatial complexity of that emission, especially within the central portion of NGC 2071-IR, and the large number of protostars as potential outflow sources within the same region, make this a challenging task.  As pointed out by E00, various mechanisms and effects can complicate the spatial distribution of the \H2\ line emission. These include precession of jets, deflection or blocking of part of an outflow by dense clumps within the flow, density non-uniformities in the ambient cloud in regions where the flow strikes the cloud, as well as the likelihood of multiple outflows possibly overlapping on the plane of the sky. In addition, the lack of detailed spectra of all of the candidates makes it difficult to identify which of them are producing winds, which have yet to become active wind sources, and which might be past their active phases.  For a few of the sources, the presence and orientations of associated masers and radio continuum emission provide invaluable information on the presence of disks and of ionized gas and on the likely orientations of outflows.

The photometry reported here suggests that while all of the objects in Table 1 lie within the NGC 2071 molecular cloud, IRS 1, IRS 2A, IRS 3, IRS 7, IRS 8, and VLA-1, each with $H - K > 2$ mag  are located in dense portions of the cloud and/or are embedded in circumstellar material. The colors of IRS 2B, IRS 4 and IRS 6A ($1 < H-K < 2$ mag) also could be due to the above factors. To what extent the red colors of these objects are due to extinction or due to thermal emission from heated circumstellar dust is not known. 

In the following subsections we briefly discuss each point source and whether direct or circumstantial evidence exists connecting it to any of the extended line emission from shocked \H2.

\subsection{ IRS 1}

In the $K$ band IRS 1 is the second brightest source in NGC 2071-IR, after IRS 6A. However, it is clear from the 10 $\mu$m map of \citet{per81} that it is by far the most luminous object in the region. TRR09 has estimated its mass to be 5 $M_{\odot}$, considerably greater than their estimate for IRS 3, and \citet{ski09} classify it as a class I protostar. One may safely conclude that it provides the bulk of the total luminosity of NGC 2071-IR, 520 $L_{\odot}$, estimated by \citet{but90}. As discussed earlier, a number of millimeter and radio observations have detected the molecular disk surrounding IRS 1 and the ionized jets being ejected from it. The orientation of the east-directed jet is shown in Figure 3 and the orientations of the disk and both jets are shown in Figure 4. They indicate that, while IRS 1 is an outflow source, it is not the one responsible for the largest outflow feature on the plane of the sky, the NE$-$SW-oriented lobes of CO and \H2\ line emission, labeled IA  and IB in Figure 3. Although it is located between the emission lobes, it is several arcseconds to the east of the axis of symmetry of that flow, as can be seen by a close examination of  Figure 3. Its location and the direction of its east-pointing jet are more consistent with its outflow being responsible for the very bright extended \H2\ line emission to its east, labeled IIA. 

If IRS 1 is the cause of the \H2\ line emission in this eastern ``lobe," then one must consider (1) why the \H2\ line emission on its opposite side (IIB in Figure 3 is so faint, and (2) why the dimensions on the sky of the outflow (coming from this most luminous protostar in NGC 2071-IR) are so small compared to those of the NE$-$SW flow. We propose that these differences are largely related to the orientation of its bipolar outflow. For example, if the true length of Outflow IIA, which extends on the sky for $\sim$0\farcm5, is comparable to the observed $\sim3'$ (E00, 0.35 pc) extent of Outflow IA, which one may assume is fairly close to being in the plane of the sky, Outflow IIA would lie 80\deg\ out of the plane of the sky.  
We suggest that shocked \H2\  to the east of IRS 1, Outflow IIA, is approaching us, such that the observer is almost looking down its barrel, and the \H2\ line emission is encountering relatively low extinction from the ambient foreground cloud. To the west the outflow from IRS 1 would be shocking \H2\ that is much deeper in the cloud and the resulting line emission would suffer much higher extinction and be much fainter, as is observed. A high-resolution spectrum of the 4.7 $\mu$m \H2\ $0-0$ $S$(9) line at IRS 1 \citep{wal05} is consistent with this interpretation; it reveals a broad line profile that peaks at -20 km s$^{-1}$ (LSR) and extends as far negative as $-50$ km s$^{-1}$. An additional test would be to compare velocity-resolved spectra of the \H2\ $S$(1) line  to the east and west of IRS 1. We note that roughly orthogonal outflows generated by closely spaced sources have been seen previously \citep{ave90,ang91}; in this case both outflows have similar dimensions on the plane of the sky.

The \H2\ line emission near IRS 1 is strongly peaked on the object, indicating that it is physically associated with the protostar. The high visual extinction ($\sim$37 mag) to the \H2\ line-emitting region thus indicates a similar high extinction to IRS 1, or possibly a higher extinction, since the \H2\ line emission probably occurs outside of a high density disk that would provide additional obscuration of the protostellar continuum. For $A_{V}$ = 37 mag, the reddening between 2.00 $\mu$m and 2.40 $\mu$m is 0.93 mag (a factor of 2.4, assuming an extinction law with a standard $\lambda^{-1.7}$ wavelength dependence), whereas for an unobscured star whose $K$-band continuum is on the Rayleigh-Jeans tail of the blackbody function, the flux density at 2.40 $\mu$m is 0.50 that at 2.00 $\mu$m. Thus, for the above extinction the observed continuum would rise only by 20\%\ across the $K$ band, whereas in Figure 5 the continuum of IRS 1 rises by 75\%.  How much of this additional rise is due to additional extinction and how much is due to thermal emission from a circumstellar disk is not known.  

\subsection{IRS 2A and IRS 2B}

Although their separations are similar, the orientation of the components of the infrared 1\farcs2 binary differ somewhat from the radio and millimeter observations of CG12, with the infrared binary oriented at position angle 108\deg\ (A to B) while the radio/millimeter binary is oriented at 120\deg. Orbital motion is ruled out by the large separation (500 a.u.); even if their masses are each 1 M$_{\odot}$) the orbital period would be $\sim$50,000 yr and the change in the orientation between the  radio and infrared observations, which are separated by 10 years, would be of order 0\farcs001. The difference in orientation could be explained by a third companion either north of IRS 2A by 0\farcs3 or south of IRS 2B by 0\farcs3.   However, neither the infrared nor the radio/images provide evidence for such an object. 

IRS 2A has considerably redder near-infrared colors than IRS 2B and is an order of magnitude brighter at $K$. At millimeter and centimeter wavelengths the two components have comparable flux densities. The centimeter-wavelength emission from each of them was interpreted by CG12 to be optically thin free-free emission, suggesting that both sources are either accreting gas or are ejecting gas into the surrounding cloud. However, the infrared composite IRS 2 spectrum in W91, obtained in a $5''$ diameter aperture, shows no signs of protostellar activity apart from \H2\ emission lines, which may not be physically associated with either source. \citet{wal05} found rather uniform emission and velocity profiles of the $0-0$ $S$(9) line across IRS 2, suggesting that the bulk of the \H2\ line emission is not associated with IRS 2.  In Figure 3 it can be seen that the binary is located both on the northern edge of Outflow IIA and the eastern edge of Outflow IA. Much more sensitive near-infrared spectra of IRS 2A and IRS 2B are easily achievable and could help determine if either of them is associated with an outflow and some of the \H2\ line emission is circumstellar. 

\subsection{IRS 3}

Based on radio data, E00 argued that IRS 3 is the most likely source of the NE$-$SW outflow in NGC 2071-IR. Examination of Figure 3 and Figure 4 shows that IRS 3 not only lies between the two lobes of the outflow, but also is close to being on its axis of symmetry, which is the most likely position for the the outflow source. In view of its much lower luminosity and lower mass (estimated by TRR09 to be $\sim$1 M$_{\odot}$) than IRS 1, it may seem surprising that it has produced such an extensive outflow ($\sim$0.6 pc on the plane of the sky from one end to the other), especially compared with the one linked to IRS 1.  However, E00 and others have found several examples of parsec-scale outflows from moderately low mass protostars. In addition, as argued previously, our view of the EW outflow, which we and E00 ascribe to IRS 1, may be considerably foreshortened. 

Closer comparison of the infrared and radio data reveals a possible problem with E00's assignment of the NE-SW outflow to IRS 3, which is illustrated in Figure 3. The position angle of the NE outflow (IA) on the sky is 45\deg (see, e.g. Figure 3 of E00), whereas the average position angle of the northern radio jet from IRS 3, measured by T98, S02, TRR09, and CG12 is 12\deg\ (indicated in the figure by the arrow), with dispersion between the measurements of only a few degrees. There is similar misalignment of the SW outflow (IB) with the southern radio jet of IRS 3. The orientation of the protostellar disk, as determined from the spatial distribution of H$_2$O masers observed by T98 and TRR09, is approximately perpendicular to the jets, as expected. In Figure 3 it can be seen that the northern radio jet lines up well with the nearby western edge of the NE outflow (IA). We note that CG12 has suggested that the IRS 3 jet is precessing, because the most recent measured orientation of the jet is position angle 15\deg.  Precession over a large range of jet position angles, from $\sim$10\deg\ to $\sim$60\deg, could account for the lateral extents of outflows IA and IB. 

The broad CO band heads in the spectrum of IRS 3 imply vigorous kinematics in the dense gas immediately surrounding the protostar. Whether this is due to Keplerian rotation of hot, dense molecular gas in a disk or emission at base of a high-velocity outflow could possibly be determined by a much higher-resolution spectrum of the band heads \citep[e.g.,][]{car93, cha93}. Assuming that the IA and IB outflows originate at IRS 3 and that those outflows lie close to the plane of the sky, with the protostellar disk roughly perpendicular, rotational motion of gas in the disk would be in the line of sight and observable as a Doppler shift. If the observed widths of the CO band heads are due to Keplerian rotational motion of, say, 150 km s$^{-1}$ (consistent with the observed width) and the mass of the protostar is 1 \Msun, the emitting gas would be 0.004 au from the center of the protostar. For a protostar the size of the sun, this would put the emission close to the surface an expected location in view of the high CO temperature. In summary, the scenario described above, in which IRS 3 is solely responsible for the NE$-$SW outflow, makes overall sense if the precession of the IRS 3 jet covers the above range of position angles.

Both the steeply rising $K$-band continuum of IRS 3 and the great strength of its $1-0$ $Q$-branch lines at the long wavelength edge of the $K$ window relative to the $1-0$ $S$-branch lines near the short wavelength edge of the window (as discussed in Section 3.3) argue for very high extinction to IRS 3 and show that it is much more deeply embedded in the NGC 2071 cloud and/or its natal material than IRS 1. The difference in visual extinction of these two protostars to their \H2\ line-emitting regions is $\sim$40 mag. The two have comparable radio fluxes (TRR09), but in view of this additional extinction, the  non-detection of Br$\gamma$ line emission from IRS 3 (the upper limit to its flux is $\sim$40 times less than IRS 1, see Figure 5) is not surprising.  As we argued previously in the case of IRS 1, the protostar itself may be even more highly obscured than its associated \H2\ line emission. The continuum flux density of IRS 3 rises by more than an order of magnitude from 2.00 $\mu$m to 2.40 $\mu$m. Such a rapid increase probably cannot be attributed to extinction alone, as that would imply $A_V$  $\sim$ 130 mag, making IRS 3 far too luminous for a 1 \Msun\ protostar at a distance of  390 pc. Thus, it is certain that a large part of the $K$-band continuum of IRS 3 is due to emission from warm circumstellar dust.

\subsection{IRS 4 and IRS 6A and B}

The $K$-band continua of IRS 4 and IRS 6 (W91) are much flatter than those of IRS 1, IRS 2, and IRS 3, suggesting, along with the photometry presented here, that neither IRS 4 nor IRS 6 is as deeply embedded in circumstellar material and/or the NGC 2071 molecular cloud as those protostars. IRS 4 is located on the same sightline as a large patch of moderately strong \H2\ line emission, as can be seen in Figure 3, and its spectrum in W91 contains prominent $1-0$ $S$(1) and $S$(0) lines. However, the location of IRS 4, $20''$ from the core of NGC 2071-IR and within Outflow IA, suggests that is not the generator of any of the outflows identified by E00.

The present observations show that IRS 6 is a binary with a separation of 1\farcs9. Its combined spectrum in W91 is similar to that of IRS 4, except that its CO absorption is much more prominent and its \H2\ line emission is much weaker; the latter is consistent with the \H2\ $S$(1) image (Figure 3). Its spectrum in W91 appears to contain a weak Br $\gamma$ emission line, but no radio detection of ionized gas has been reported. Because in the $K$ band IRS 6A is three times brighter than IRS 6B, it is likely that the observed strong CO absorption observed by W91 arises in its photosphere. Like IRS 4, the nondescript location of IRS 6A/B in relation to the outflows IA-III suggests that it is not likely to be the driver of any of them. 

\subsection{IRS 7}

The $K$-band spectrum of IRS 7 in W91, contains strong CO overtone band emission, evidence that IRS 7 is an accreting protostar that could therefore also be an outflow source. E00 suggested that IRS 7 is responsible for Outflow III, which extends from the protostar to the northwest, possibly as far or farther than the northwest edge of Figure 3.  There appears to be a knotty arc or bow along the northwest edge of the nearest patch of \H2\ line emission, which points back to IRS 7. Like the spectrum of IRS 3, the one of IRS 7 in W91 does not show Br $\gamma$ emission. If the blobs of shocked \H2\ extending to the northwest are one side of bipolar outflow from IRS 7, as suggested by E00, the other lobe would intersect flows IA and IIA and contribute to their confusing \H2\ morphologies. 

\subsection{IRS 8}

IRS 8 is one of the redder point sources in NGC 2071-IR (see Table 1). Its high polarization (W93) suggests that it is embedded in its own dusty cocoon. Its location, only $10''$ southwest of IRS 1 and IRS 3 and adjacent to a bright clump of \H2\ line emission to its northeast, might be a sign that it is the source of another outflow. However, the \H2\ $0-0$ $S$(9) line near its position \citep{wal05} is redshifted and is consistent with the line emission arising in the SW outflow (IB). No near-infrared spectrum of IRS 8 exists in the literature, nor have radio or millimeter observations of it been reported. Thus, any suggestions as to its nature and to its physical relationship to the outflows in NGC 2071-IR would be highly speculative at present. 

\subsection{VLA-1}

This compact 1.3 and 3.6 cm radio continuum source (TRR09, S09) has been detected in the near-infrared only in the $K$ band, and  is the faintest of the all of the point sources in NGC 2071-IR in that band (Table 1). We concur with TRR09 that VLA-1 is probably at an earlier evolutionary stage than IRS 1 and probably the other point sources as well. Its location in the gap between the NE and SW lobes (outflows IA and IB) and apparently closer to the axis of symmetry of those two outflows is intriguing. The radio continuum emission from VLA-1 has been interpreted by TRR09 as originating in a hyper-compact or ultra-compact H\two\ region, a possible indication that the source is already ejecting material. A $K$-band spectrum of this object could give important clues as to its nature. 

\section{Present Understanding and Future Observations}

The near-infrared images of NGC 2071-IR presented in this paper, and in particular the image of the \H2\  $1-0$ $S$(1) emission line, have given a highly detailed view of this complex region, which apparently contains multiple molecular outflows from embedded protostars. However, clearly identifying individual outflows and assigning individual protostars to them remain problematical. Much could be learned about the natures, activities, and evolutionary states of infrared point sources in NGC 2071-IR from future infrared spectroscopy. This is illustrated by the medium-resolution $K$-band spectra of IRS 1 and IRS 3 reported here, which attest to each of them being in an active phases of accretion and/or outflow. Higher-resolution spectra of selected \H2\ and H\one\ lines, and overtone CO bands in them would provide critical details pertaining to the distributions of their emitting gas and their kinematics.  Similar medium- and high-resolution spectra of the remaining ``mystery" objects within this region, IRS 2, 4, 6, 7, 8 and VLA-1, could determine whether any of them are likewise active. In addition, occasional monitoring of the radio jets in IRS 1 and IRS 3 could clarify their relationships to the outflows proposed by E00 and by us to be associated with them.

\begin{acknowledgements}
 
This paper is based on observations obtained at the Gemini Observatory, which is operated by the Association of Universities for Research in Astronomy, Inc., under a cooperative agreement with the NSF on behalf of the Gemini partnership: the National Science Foundation (United States), the National Research Council (Canada), CONICYT (Chile), Ministerio de Ciencia, Tecnologia e Innovacion Productiva (Argentina), Ministerio da Ciencia, Tecnologia e Inovacao (Brazil), and Korea Astronomy and Space Science Institute (Republic of Korea).  We thank the staff of Gemini for its support and the referee for a number of helpful comments.

\end{acknowledgements}

\end{document}